\documentclass[numberedheadings]{aipproc}
\layoutstyle{6x9}

\begin{document}

\title{Synchronised neutrino oscillations from self interaction and associated applications}
\author{Yvonne Y. Y. Wong}{address={Department of Physics and Astronomy,
 University of Delaware, Newark, DE 19716}}

\begin{abstract}
A recent revival of interest in synchronised oscillations due to
neutrino--neutrino forward scattering in dense gases has led to
two interesting applications with notable outcomes: (i)
cosmological bounds on neutrino--antineutrino asymmetries are
improved owing to flavour equilibration prior to the onset of big
bang nucleosynthesis, and (ii) a neutron-rich environment required
for $r$-process nucleosynthesis is shown to be always maintained
in a supernova hot bubble irrespective of flavour oscillations,
contrary to results from previous studies.  I present in this talk
a pedagogical review of these works.
\end{abstract}

\maketitle

\section{The neutrino--neutrino  forward scattering saga}
\label{sec:section1}

It is well known that neutrino oscillations in a medium are
affected by the presence of other particles.  A familiar example
is oscillations in the sun:  Extra interaction channels
 available exclusively to the electron neutrino
induce for it an excess ``effective'' mass  through coherent
forward scattering on the ambient electrons
\cite{bib:wolfenstein}.\footnote{This statement assumes that
$\nu_e$ oscillates into $\nu_{\mu}$ and/or $\nu_{\tau}$.  If a
sterile neutrino is involved, then neutral current scattering of
$\nu_e$ on the surrounding electrons and nucleons will also
contribute to the excess \cite{bib:bargersterile}.  Note that we
shall not be considering sterile neutrinos in this talk.}
Depending on the electron number density, this excess serves as a
refractive index for the oscillating neutrino system,  modifies
the oscillation frequency and amplitude from their vacuum values,
and is responsible for such phenomenon as the
Mikheyev--Smirnov--Wolfenstein (MSW) effect \cite{bib:msw} and,
with it, the large mixing angle (LMA) solution to the solar
neutrino problem \cite{bib:solarfits}.

In certain astrophysical and cosmological settings, neutrinos form
a gas so dense that forward scattering of neutrinos on
the neutrinos themselves may constitute a significant source of
refraction \cite{bib:notzold,bib:mayle}. Furthermore, such scattering differs
markedly from scattering on other fermions, since in the former
instance it is possible for the ``beam'' and ``background''
neutrinos to exchange flavour via the neutral current.  The net
outcome is a set of ``on-''  and ``off-diagonal'' refractive
indices that are dependent on the state of every other neutrino in
the gas, thereby rendering the evolution of the whole ensemble
highly nonlinear.

In the very earliest works incorporating the refractive index from
neutrino--neutrino forward scattering, the off-diagonal refractive
indices were erroneously left out. This was rectified by
Pantaleone in 1992 \cite{bib:pantaleone}.  Shortly after, the
implications of these ``self interactions'' for a multi-momentum
gas were investigated extensively by
Samuel \cite{bib:samuel}, who discovered in his numerical studies
that when a certain
critical density is reached, the gas begins to exhibit
self-maintained coherence (a.k.a. synchronisation)
when decoherence is {\it a priori} expected. This interesting
effect went unexplained for almost a decade, and was recently
reexamined by Pastor, Raffelt, and Semikoz \cite{bib:synch}, who
succeeded in giving it a clean and physical interpretation on
which I shall elaborate  in this talk.

The result of Ref.\ \cite{bib:synch} has since been extended and
applied to the only two environments in which self interactions
are expected to play a dominant role: the early universe, and a
core collapse supernova.  In the first setting, synchronised
oscillations are investigated in the context of relic
neutrino--antineutrino asymmetries in the epoch preceding big bang
nucleosynthesis (BBN). Independent studies by several groups have
demonstrated that if the neutrino oscillation parameters are
indeed those inferred from the solar and atmospheric neutrino
data, then the asymmetries in the $\nu_{\mu}$ and the $\nu_{\tau}$
sectors must be equilibrated with that in the $\nu_e$  sector
prior to BBN, and be subject to the same stringent constraints on
the $\nu_e$ asymmetry derived from uncertainties in the primordial
$^4{\rm He}$ abundance \cite{bib:dhpprs,bib:yyyw,bib:kev} (see
also Ref.\ \cite{bib:lunardini}). In the second case, self interactions
modify in an unexpected way the flavour evolution of both
neutrinos and antineutrinos in the hot bubble region above a
nascent neutron star a few seconds after core bounce
\cite{bib:sn}. This impinges directly on the success or otherwise
of $r$-process nucleosynthesis, which is believed to take place in
this setting.

In this talk, I shall present a pedagogical review of these works,
starting with the interpretation of synchronised oscillations in
Sec.\ \ref{sec:section1}, and then moving on to the applications in
Secs.\ \ref{sec:section2}
and \ref{sec:section3}.
The material used here is drawn largely from Refs.\
\cite{bib:synch}, \cite{bib:dhpprs}, \cite{bib:yyyw}, \cite{bib:kev},
 and \cite{bib:sn}. If you the
reader happen to the author of one of these works, please pardon me
for quoting you almost verbatim in some places.

\subsection{Equations of motion}
\subsubsection{Density matrix formalism}

We begin with a simple two-flavour system.  In order to understand
the synchronisation phenomenon, it is convenient to work with
bilinears of the neutrino wave function (and hence density
matrices), since, as we shall see, the neutrino self
interaction potential appears also in this form.

For a group of neutrinos with momentum $p$, the density matrix at
time $t$ in the flavour basis is defined as
\begin{eqnarray}
\rho(p,t) & \equiv & \left( \begin{array}{cc}
                    \rho_{\alpha \alpha} & \rho_{\alpha \beta} \\
                    \rho_{\beta \alpha} & \rho_{\beta \beta}
                    \end{array} \right)
                    \doteq   \frac{1}{N_0 (p)}
                    \sum_i  | \psi_i (p,t)\rangle \langle \psi_i
                    (p,t)| \nonumber \\
                    & = & \frac{1}{N_0 (p)} \sum_i \left(\begin{array}{cc}
                     |a_i(p,t)|^2 & a_i (p,t) b_i^*(p,t) \\
                 a_i^*(p,t) b_i(p,t) &  |b_i(p,t)|^2
                    \end{array} \right),
\end{eqnarray}
where the subscripts $\alpha$ and $\beta$ denote two different
neutrino flavours, $N_0 (p)$ is some time-independent
normalisation factor, and
\begin{equation}
|\psi_i (p,t) \rangle = a_i(p,t) |\nu_{\alpha} \rangle + b_i(p,t)
|\nu_{\beta} \rangle
\end{equation}
is the wave function of the $i$th neutrino in the group.  The
functions $|a_i(p,t)|^2$ and   $|b_i(p,t)|^2$ are probabilities in
the one-particle formulation.  Thus the sums
\begin{equation} N_{\nu_{\alpha}} \equiv \sum_i |a_i(p,t)|^2 = \rho_{\alpha \alpha} N_0(p), \qquad
N_{\nu_{\beta}} \equiv \sum_i |b_i(p,t)|^2=\rho_{\beta \beta}
N_0(p)
\end{equation}
in the diagonal of $\rho (p,t)$ represent respectively the
occupation numbers of $\nu_{\alpha}$ and $\nu_{\beta}$ at a given
momentum, while the off-diagonal $\rho_{\alpha \beta}$ and
$\rho_{\beta \alpha}$ are intangible quantities containing the
system's  phase information.

 We evolve $\rho (p,t)$ by taking its
commutator with the Hamiltonian $H(p,t)$ in flavour space,
\begin{equation}
\label{eq:eqofmotion}
\partial_t \rho(p,t) = - i [H (p,t),\rho(p,t)].
\end{equation}
As an example, the Hamiltonian governing $\nu_e \leftrightarrow
\nu_{\mu,\tau}$ oscillations in the sun takes the form
\begin{equation}
H(p,t)  = \frac{\delta m^2}{4p}
    \left(\begin{array}{cc}
            -\cos 2 \theta & \sin 2 \theta \\
            \sin 2 \theta & \cos 2 \theta \end{array} \right) +
    \left( \begin{array}{cc}
           A(p,t) & 0 \\
            0 & 0 \end{array} \right),
\end{equation}
in which the first term describes vacuum oscillations, and the
second term with matter potential $ A(p,t)=\sqrt{2} G_F n_{e^-}
(t)$, where $n_{e^-}(t)$ denotes the solar electron number
density, is responsible for refractive effects.\footnote{Rigorous
derivations of the equation of motion and various interaction
potentials can be found in Refs.\ \cite{bib:mck&t} and
\cite{bib:sigl}.  See
also Ref.\ \cite{bib:stodolsky}.}

For our purposes,  it is useful to recast the equation of motion
(\ref{eq:eqofmotion}) into  Bloch form by first parameterising
$\rho(p,t)$ and $H(p,t)$ in terms of the functions $(P_0,\ {\bf
P})$ and $(V_0, \ {\bf V})$:
\begin{eqnarray}
\rho (p,t) \!\!& = & \!\! \frac{1}{2} [P_0  + {\bf P}  \cdot {\bf
\sigma}]\  = \ \frac{1}{2} \left( \begin{array}{cc}
                    P_0 + P_z & P_x - i P_y \\
                    P_x + iP_y & P_0 - P_z \end{array} \right),
                    \nonumber \\
 H (p,t) \!\! & = & \!\! \frac{1}{2} [V_0 + {\bf V} \cdot {\bf \sigma}]
\ = \  \frac{1}{2} \left( \begin{array}{cc}
                    V_0 + V_z & V_x - i V_y \\
                    V_x + i V_y & V_0 - V_z \end{array} \right).
\end{eqnarray}
Here, ${\bf \sigma} = \sigma_x {\bf x} + \sigma_y {\bf y}+
\sigma_z {\bf z} $ are the Pauli matrices.  The quantities ${\bf
P}=P_x {\bf x} + P_y {\bf y} + P_z {\bf z}$ and ${\bf V}= V_x {\bf
x} + V_y {\bf y} + V_z {\bf z}$ may be regarded, respectively, as
a three-dimensional ``spin polarisation'' and a ``magnetic field''
vector, whose time and momentum dependences are implicit. The
equation of motion (\ref{eq:eqofmotion}) transforms into
\begin{equation}
\label{eq:bloch}
\partial_t {\bf P}  =  {\bf V} \times {\bf P}, \qquad \partial_t P_0 =
0,
\end{equation}
such that the dynamics of the oscillating system is analogous to a
spin vector predisposed to precessing around a magnetic field
${\bf V}$ at a rate $|{\bf V}|$, with
\begin{equation}
P_z  =  \frac{N_{\nu_{\alpha}}(p,t) -
N_{\nu_{\beta}}(p,t)}{N_0(p)}
\end{equation}
parameterising the excess of $\nu_{\alpha}$ over $\nu_{\beta}$,
and conservation of probability demands that the function
\begin{equation}
P_0 = \frac{N_{\nu_{\alpha}}(p,t) + N_{\nu_{\beta}}(p,t)}{N_0(p)}
\end{equation}
remain constant with time.

Observe in Eq.\ (\ref{eq:bloch}) that  $V_0$ is
redundant for flavour oscillations, since it contributes only an
unobservable overall phase to the neutrino wave function.  Some
typical examples of the field vector ${\bf V}$ are (i)
\begin{equation}
{\bf V} = \frac{\delta m^2}{2p} {\bf B} = \frac{\delta m^2}{2p} (
\sin 2 \theta {\bf x}  - \cos 2 \theta {\bf z}),
\end{equation}
for oscillations in vacuum, where, pictorially, the unit vector
${\bf B}$ forms an angle $2 \theta$ with the negative $z$-axis and
$\delta m^2/2p$ is the precession frequency, and (ii)
\begin{equation}
{\bf V} = \frac{\delta m^2}{2p} {\bf B} + A(p,t) {\bf z}=
\frac{\delta m^2}{2p} \sin 2 \theta {\bf x} + [ A(p,t) -
\frac{\delta m^2}{2p} \cos 2 \theta ] {\bf z},
\end{equation}
for oscillations in a medium, with the matter potential $A(p,t)$
entering in the $z$-direction.  As mentioned before, the presence
of $A(p,t)$ generally results in the modification of the
oscillation frequency and amplitude.  In the spin precession
picture, this is equivalent to defining a field vector ${\bf
B}_{\rm eff}$
\begin{eqnarray}
\frac{\delta m^2}{2p} {\bf B} + A(p,t) {\bf z}=
 \frac{\delta m^2_{\rm eff}}{2p} (
\sin 2 \theta_{\rm eff} {\bf x}  - \cos 2 \theta_{\rm eff} {\bf
z}) \equiv \frac{\delta m^2_{\rm eff}}{2p} {\bf B}_{\rm eff},
\end{eqnarray}
which makes an angle $2 \theta_{\rm eff}$ with the negative
$z$-axis, and ${\bf P}$ precesses around is at a rate $\delta
m^2_{\rm eff}/2p$, where
\begin{eqnarray}
\label{eq:angle}
\sin 2 \theta_{\rm eff} & =&  \frac{\sin 2
\theta}{\sqrt{( 2p A(p,t) /\delta m^2  - \cos 2 \theta)^2 + \sin^2
2 \theta}}, \nonumber
\\ \delta m^2_{\rm eff} &=&  \delta m^2
 \sqrt{(2pA(p,t)/\delta m^2  - \cos 2 \theta)^2 + \sin^2 2 \theta}.
\end{eqnarray}
If $|A(p,t)|$ is much larger than $|\delta m^2/2p|$, the former
will force the total field vector to align with the $z$-axis.  The
effective mixing angle $\theta_{\rm eff}$ approaches $0$ or $\pi/2$,
thereby suppressing flavour oscillations. On the other hand, if
the condition
\begin{equation}
\label{eq:resonancecondition} A(p,t)= \frac{\delta m^2}{2p} \cos 2
\theta
\end{equation}
happens to hold, then the vector ${\bf V}$ points in the
$x$-direction, forming a right angle with the $z$-axis which
corresponds to maximal mixing, $\theta_{\rm eff}=\pi/4$.  This
is the so-called resonance condition.

Finally, observe that the field  vector ${\bf B}_{\rm eff}$
differs for each momentum mode.  Thus the spin vectors of an
ensemble of neutrinos possessing a spectrum of momenta do not in
general precess in the same direction, or at the same rate.

\subsubsection{Add some neutrino--neutrino forward scattering}

Suppose now we have a spatially homogeneous and isotropic neutrino
gas of volume $V$.  Contribution from neutrino--neutrino forward
scattering to the total field vector ${\bf V}$ appears in the form
\cite{bib:sigl}
\begin{equation}
\frac{\sqrt{2} G_F}V {\bf J},
\end{equation}
where
\begin{equation}
{\bf J} = \sum_k {\bf P}_k N_0(p_k)
\end{equation}
is the total polarisation vector, and the subscript $k$ labels the
$k$th momentum mode. Thus the simplest equation of motion one can
write down to describe flavour oscillations in the $k$th momentum
mode of a self-interacting neutrino gas will consist of a term
from vacuum oscillations, and another from  self
interactions:
\begin{equation}
\label{eq:singlemode}
\partial_t {\bf P}_k  = \frac{\delta m^2}{2p_k} {\bf B} \times {\bf
P}_k + \frac{\sqrt{2} G_F}{V} {\bf J} \times {\bf P}_k.
\end{equation}
Note that Eq.\ (\ref{eq:singlemode}) is valid {\it only} for the case
of active--active flavour oscillations.  For an active--sterile
system, {\it all} interaction potentials must couple exclusively
to the active neutrino, and their contributions to the total field
vector ${\bf V}$ in the Bloch formulation always point in the $\pm
z$-direction.

\subsection{Physical interpretation of synchronised oscillations}

\subsubsection{Vacuum-like oscillations}

In the pioneering work of Samuel \cite{bib:samuel}, Eq.\
(\ref{eq:singlemode}) was solved numerically for a neutrino gas
with a distribution of momenta.   To his surprise, the oscillating
multi-momentum gas exhibited synchronised, monochromatic
behaviours when the self interaction potential was allowed to
dominate over the vacuum oscillation rate:
\begin{equation}
\label{eq:densegas} \frac{\sqrt{2} G_F}{V} |{\bf J}| \gg \left|
\frac{\delta m^2}{2p_k} \right|.
\end{equation}
A series of papers by Samuel and co-workers on the same subject
emerged in the years following this initial discovery
\cite{bib:moresamuel,bib:kostelecky,bib:stability}, extending the results of Ref.\
\cite{bib:samuel} to include antineutrinos and a non-neutrino
background in the context of the early universe. Exact solutions
to the equations of motion were found \cite{bib:kostelecky}, and
the stability of the synchronised state analysed
\cite{bib:stability}. Nonetheless, a clean, physical explanation
for the synchronisation phenomenon had continued to elude the
players. This interesting effect was elucidated recently by
Pastor, Raffelt, and Semikoz in Ref.\ \cite{bib:synch}, whose
argument goes as follows.

When the neutrino gas is sufficiently dense and the condition
(\ref{eq:densegas}) satisfied,  the equation of motion
(\ref{eq:singlemode}) for a single momentum mode becomes dominated
on the right hand side by the self interaction
potential:
\begin{equation}
\partial_t {\bf P}_k \simeq \frac{\sqrt{2} G_F}{V} {\bf J}
\times {\bf P}_k.
\end{equation}
This implies an approximate solution
\begin{equation}
\label{eq:approxsolution} {\bf P}_k \simeq ({\bf P}^{\rm i}_k
\cdot \hat{\bf J}^{\rm i} ) \hat{\bf J} +\  {\rm precessions} \;
\, {\rm around} \; \, {\bf J},
\end{equation}
where the superscript ``i'' denotes initial.  In plain English,
all polarisation vectors ${\bf P}_k$ across the momentum distribution
are individually first and foremost
drawn towards the giant total polarisation vector ${\bf J}$, and
precess around it at a rapid rate of $\sim \sqrt{2} G_F |{\bf
J}|/{V}$. If the evolution of ${\bf J}$ proceeds at a rate much
slower than this precession frequency, then  all ${\bf P}_k$'s
 will remain pinned to ${\bf J}$, which, as we shall see, forms the
basis of synchronisation.

The dynamics of ${\bf J}$ is governed by an equation of motion
constructed from summing the single mode equation
(\ref{eq:singlemode}) over all momentum modes,
\begin{equation}
\partial_t {\bf J} = {\bf B} \times \sum_k \frac{\delta m^2}{2
p_k} {\bf P}_k N_0(p_k).
\end{equation}
Substituting the approximate solution (\ref{eq:approxsolution})
and averaging  the oscillations to zero over the momentum spread, we
immediately see that the evolution of ${\bf J}$ is
non-dissipative,
\begin{equation}
\label{eq:approxj}
\partial_t {\bf J} \simeq \left\langle \frac{\delta m^2}{2p}
\right\rangle {\bf B} \times {\bf J},
\end{equation}
with
\begin{equation}
\left\langle \frac{\delta m^2}{2p} \right\rangle \equiv
\frac{\sum_k \frac{\delta m^2}{2 p_k} ({\bf P}_k^{\rm i} \cdot
\hat{\bf J}^{\rm i} )N_0(p_k)}{| \sum_k {\bf P}_k N_0(p_k) |}.
\end{equation}
Since the slowly-moving part of every polarisation vector ${\bf
P}_k$  follows to first approximation the dynamics of ${\bf J}$,
an apparently monochromatic behaviour emerges within the
multi-momentum neutrino gas, and $\left\langle \frac{\delta
m^2}{2p} \right\rangle$ is the synchronised oscillation frequency
with an explicit dependence on the initial conditions.

\subsubsection{Add a non-neutrino background}

Granted that neutrino--neutrino forward scattering continues to
constitute
the dominant interaction potential, it is easy to generalise the
previous picture to include a non-neutrino background $A(p,t)$.
The equation of motion is essentially the same as Eq.\
(\ref{eq:singlemode}), save for the replacement
\begin{equation}
\frac{\delta m^2}{2p_k} {\bf B} \to  \frac{\delta m^2}{2p_k} {\bf
B} + A(p_k,t) {\bf z},
\end{equation}
and the analogue of the ``dense gas'' condition
(\ref{eq:densegas}) is
\begin{equation}
\frac{\sqrt{2} G_F}{V} |{\bf J}| \gg \left| \frac{\delta
m^2}{2p_k} \right|, \, |A(p_k,t)|.
\end{equation}
When this condition is satisfied, all polarisation vectors ${\bf
P}_k$ become pinned to ${\bf J}$ as per the approximate solution
(\ref{eq:approxsolution}), and the evolution of ${\bf J}$ is
equally well described by Eq.\ (\ref{eq:approxj}) upon replacing
\begin{equation}
\label{eq:effectivev}
\left\langle \frac{\delta m^2}{2p}
\right\rangle {\bf B} \to \left\langle \frac{\delta m^2}{2p}
\right\rangle {\bf B} + \left\langle A(p,t) \right\rangle {\bf z},
\end{equation}
where
\begin{equation}
\left\langle A(p,t) \right\rangle \equiv \frac{\sum_k A(p_k,t)
({\bf P}_k^{\rm i} \cdot \hat{\bf J}^{\rm i} ) N_0(p_k)}{| \sum_k
{\bf P}_k N_0(p_k)|}
\end{equation}
represents an average background.

Perhaps the most interesting aspect of this result is that a
strong self interaction potential not only forces all modes in a
multi-momentum neutrino gas to oscillate at the same frequency,
but also with the same amplitude, regardless of the individual
modes' response to the non-neutrino background.  Previously, we
saw that electrons in the solar interior alter the mixing angle
and the oscillation frequency for $\nu_e \leftrightarrow \nu_{\mu,
\tau}$ oscillations in a momentum-dependent fashion.  In the
present case, however, the effective field vector for ${\bf J}$ in
Eq.\ (\ref{eq:effectivev}) necessarily implies a common mixing
angle
\begin{equation}
\label{eq:synchangle}
\sin 2 \theta_{\rm synch}  =  \frac{\sin 2
\theta}{\sqrt{(\langle A(p,t) \rangle/\langle \frac{\delta
m^2}{2p} \rangle - \cos 2 \theta)^2 + \sin^2 2 \theta}},
\end{equation}
and a common oscillation frequency
\begin{equation}
\label{eq:synchfreq} \omega_{\rm synch} = \left\langle
\frac{\delta m^2}{2p} \right\rangle \sqrt{\left(\langle A(p,t)
\rangle/\left\langle \frac{\delta m^2}{2p} \right\rangle - \cos 2
\theta \right)^2 + \sin^2 2 \theta},
\end{equation}
for all momentum modes.  A corollary of Eqs.\
(\ref{eq:synchangle}) and (\ref{eq:synchfreq}) is that if the
``average'' resonance condition
\begin{equation}
\langle A(p) \rangle = \left\langle \frac{\delta m^2}{2p}
\right\rangle \cos 2 \theta
\end{equation}
is met, then it is necessarily met by {\it all} momenta at the
same time or at the same place, and any MSW resonant conversion
that is incurred must proceed with the same adiabaticity across
the spectrum \cite{bib:yyyw,bib:kev}.

\subsubsection{Add some antineutrinos}

In a realistic system such as the early universe, it is inevitable
that neutrinos and antineutrinos will coexist, in which case it is
necessary to consider also neutrinos scattering on antineutrinos
and vice versa.  To this end, we first define the density matrix
for antineutrinos of momentum $p$ in flavour space, and the
corresponding polarisation vector $\bar{\bf P}$:
\begin{equation} \bar{\rho}(p) = \left( \begin{array}{cc}
                    \bar{\rho}_{\alpha \alpha} & \bar{\rho}_{\beta \alpha} \\
                    \bar{\rho}_{\alpha \beta} & \bar{\rho}_{\beta \beta}
                    \end{array} \right) =\frac{1}{2} [\bar{P}_0 +
                    \bar{{\bf P}} \cdot {\bf \sigma}].
\end{equation}
To first order in $G_F$, the equations of motion are
\cite{bib:sigl}
\begin{eqnarray}
\label{eq:anti}
\partial_t {\bf P}_k \! \!&=& \!\!+ \frac{\delta m^2}{2p_k} {\bf B}
\times {\bf P}_k + (A_{CP+} + A_{CP-})\  {\bf z} \times {\bf P}_k
+ \frac{\sqrt{2} G_F}{V} ({\bf J} - \bar{\bf J}) \times {\bf P}_k,
\nonumber \\ \partial_t \bar{\bf P}_q \!\!&=&\!\! - \frac{\delta
m^2}{2p_q} {\bf B} \times \bar{{\bf P}}_q - (A_{CP+} - A_{CP-})\
{\bf z} \times \bar{{\bf P}}_q + \frac{\sqrt{2} G_F}{V} ({\bf J} -
\bar{\bf J}) \times \bar{{\bf P}}_q,
\end{eqnarray}
where $\bar{\bf J}$ is the antineutrino analogue of ${\bf J}$, and
we have segregated the background interaction potentials into a
$CP$ symmetric ($A_{CP+}$) and a $CP$ asymmetric ($A_{CP-}$) part,
with implicit momentum and time dependences.  A $CP$ asymmetric
potential, such as that arising from a local four-fermion
coupling, carries opposite signs in the interaction Hamiltonian
for neutrinos and antineutrinos, while $A_{CP+}$ is identical for
both $\nu$ and $\bar{\nu}$, and comes usually from higher order
propagator effects \cite{bib:notzold,bib:sigl}.

Synchronisation occurs when the condition
\begin{equation}
\label{eq:synchi} \frac{\sqrt{2}G_F}{V} |{\bf J} - \bar{\bf J}|
\gg \left| \frac{\delta m^2}{2p_k} \right|, \, |A_{CP+}|,\,
|A_{CP-}|
\end{equation}
holds, and the vector ${\bf I} \equiv {\bf J} - \bar{\bf J}$ now
assumes the role played by the total polarisation vector ${\bf J}$
in the previous neutrino-only case.  All ${\bf P}_k$ and $\bar{\bf
P}_q$ become pinned to ${\bf I}$,
\begin{eqnarray}
{\bf P}_k  & \simeq & ({\bf P}_k^{\rm i} \cdot \hat{\bf I}^{\rm
i}) \hat{\bf I}  +\  {\rm precessions} \; \, {\rm around} \; \,
{\bf I}, \nonumber \\  \bar{\bf P}_q  &\simeq & (\bar{\bf
P}_q^{\rm i} \cdot \hat{\bf I}^{\rm i}) \hat{\bf I}  +\  {\rm
precessions} \;\, {\rm around} \; \, {\bf I},
\end{eqnarray}
and the corresponding evolution equation for the compound system is
\begin{equation}
\label{eq:compound}
\partial_t {\bf I} \simeq \left[ \left\langle \frac{\delta m^2}{2p}
\right\rangle {\bf B} + ( \langle A_{CP+} \rangle + \langle
A_{CP-} \rangle ) \ {\bf z} \right] \times {\bf I},
\end{equation}
where
\begin{eqnarray}
\left\langle \frac{\delta m^2}{2p} \right\rangle &=& \frac{\sum_k
\frac{\delta m^2}{2p_k} ({\bf P}^{\rm i}_k \cdot \hat{\bf I}^{\rm
i})N_0(p_k) + \sum_q \frac{\delta m^2}{2p_q} (\bar{\bf P}^{\rm
i}_q \cdot \hat{\bf I}^{\rm i})N_0(p_q)}{| \sum_k {\bf P}_k
N_0(p_k) - \sum_q \bar{\bf P}_q N_0(p_q) |}, \nonumber
\\
\left\langle A_{CP+} \right\rangle &=& \frac{\sum_k A_{CP+} ({\bf
P}^{\rm k}_i \cdot \hat{\bf I}^{\rm i})N_0(p_k) + \sum_q A_{CP+}
(\bar{\bf P}^{\rm i}_q \cdot \hat{\bf I}^{\rm i})N_0(p_q)}{|
\sum_k {\bf P}_k N_0(p_k) - \sum_q \bar{\bf P}_q N_0(p_q) |},
\nonumber\\ \left\langle A_{CP-} \right\rangle &=& \frac{\sum_k
A_{CP-} ({\bf P}^{\rm i}_k \cdot \hat{\bf I}^{\rm i})N_0(p_k) -
\sum_q A_{CP-} (\bar{\bf P}^{\rm i}_q \cdot \hat{\bf I}^{\rm
i})N_0(p_q)}{| \sum_k {\bf P}_k N_0(p_k) - \sum_q \bar{\bf P}_q
N_0(p_q) |}.
\end{eqnarray}
Observe that the present formulation is invalid if
\begin{equation}
\sum_k {\bf P}_k N_0(p_k) \neq \sum_q \bar{\bf P}_q N_0(p_q),
\end{equation}
although the case of identical $\nu$ and $\bar{\nu}$ spectra can
lead to some curious effects \cite{bib:synch}.  It is interesting
to note that both neutrinos and antineutrinos are synchronised and
evolve in an identical manner, regardless of the presence of a
background medium that distinguishes between the $CP$ partners
\cite{bib:yyyw,bib:sn}. This has an interesting implication for
$r$-process nucleosynthesis in a supernova hot bubble, to be
discussed later.

\vspace{5mm}
\begin{center}
*******************************
\end{center}

This is as much background material as we shall need.  Let us now
turn to the applications.

\section{Application 1: Equilibration of relic
neutrino--antineutrino
asymmetries}

\label{sec:section2}

One of the many open questions in cosmology is the possibility of
admitting a large relic neutrino--antineutrino asymmetry,
\begin{equation}
L_{\nu_{\alpha}} \equiv \frac{n_{\nu_{\alpha}} -
n_{\bar{\nu}_{\alpha}}}{n_{\gamma}},
\end{equation}
where $n_{\psi}= \int N_{\psi} p^2 dp$ denotes the number density
of the particle species $\psi$.  At temperatures $T
> 2 \ {\rm MeV}$, thermal and chemical equilibria are generally believed to
hold for an ensemble of neutrinos and antineutrinos of some active flavour $\alpha$.
The distribution functions $N_{\nu_{\alpha}}$ and
$N_{\bar{\nu}_{\alpha}}$ assume the Fermi--Dirac form,
\begin{equation}
N_{\nu_{\alpha}} = N_{\rm eq}( \xi_{\nu_{\alpha}}),
\qquad N_{\bar{\nu}_{\alpha}}  = N_{\rm eq}
(\xi_{\bar{\nu}_{\alpha}}),
\end{equation}
where
\begin{equation}
\label{eq:fermidirac}
 N_{\rm eq}( \xi)=   \frac{1}{2 \pi^2} \frac{p^2 dp}{1 +
e^{p/T- \xi}},
\end{equation}
and $\xi_{\bar{\nu}_{\alpha}} = - \xi_{\nu_{\alpha}}$, such that
the asymmetry in $\nu_{\alpha}$ and $\bar{\nu}_{\alpha}$ may be
alternatively expressed in terms of the species' chemical
potential $\xi_{\nu_{\alpha}}$,
\begin{equation}
L_{\nu_{\alpha}} = \frac{1}{n_{\gamma}} \int (N_{\nu_{\alpha}} -
N_{\bar{\nu}_{\alpha}})= \frac{1}{12 \zeta (3)} ( \pi^2
\xi_{\nu_{\alpha}} + \xi_{\nu_{\alpha}}^3),
\end{equation}
with $n_{\gamma} = 2 \zeta(3) T^3/\pi^2$, and $\zeta$ is the
Riemann zeta function.\footnote{The terms neutrino--antineutrino
asymmetry, neutrino chemical potential, and neutrino degeneracy
parameter will be used interchangeably throughout this talk.}

As of now, there are no direct observations of the cosmic neutrino
background. Thus the existence or otherwise of any sizeable $\xi$
can only be established indirectly from the study of the cosmic
microwave background radiation (CMBR) anisotropy spectrum, and
from requiring consistency with the observationally highly
successful theory of BBN. In the former
case, we note that the contribution by one neutrino species
($\nu_{\alpha}$ and $\bar{\nu}_{\alpha}$) to the total energy
density in radiation in the universe depends immediately on the
species' chemical potential:
\begin{equation}
\label{eq:energydensity}
\rho_{\nu_{\alpha} \bar{\nu}_{\alpha}} =
T^4 \frac{7 \pi^2}{120} \left[ 1 + \frac{30}{7} \left(
\frac{\xi_{\nu_{\alpha}}}{\pi} \right)^2 + \frac{15}{7} \left(
\frac{\xi_{\nu_{\alpha}}}{\pi} \right)^4 \right].
\end{equation}
An increase in the energy density in radiation tends to delay the
epoch of matter--radiation equality, which is known to magnify the
amplitude of the first acoustic peak in the CMBR angular power
spectrum, and shift all other peaks to higher multipoles
\cite{bib:les}. Currently, CMBR limits $\xi$ to below $3$,
applicable to all neutrino flavours \cite{bib:hannestad}.

Secondly, an asymmetry in the electron neutrino sector impinges
directly on the $\beta$-processes
\begin{equation}
\label{eq:betaprocesses} n + \nu_e  \rightleftharpoons  p + e^-,
\qquad  p+ \bar{\nu}_e  \rightleftharpoons  n + e^+
\end{equation}
that are responsible for setting the neutron-to-proton ratio prior
to BBN ($T \sim 1 \ {\rm MeV}$)---a positive $\xi_{\nu_e}$ tends
to lower $n_n/n_p$ while a negative $\xi_{\nu_e}$ raises it. Since
the primordial $^4{\rm He}$ mass fraction $Y_p$ depends
exclusively on the value of $n_n/n_p$ at the weak freeze-out,
 the measurement of $Y_p$ can be used to limit
the allowed values of $\xi_{\nu_e}$, which cover the range $-0.01 <
\xi_{\nu_e} < 0.07$ at present \cite{bib:dhpprs,bib:dbbn}.
Furthermore, a nonzero $\xi$ in any neutrino flavour tends to
accelerate the expansion of the universe through its contribution
to the energy density {\it a la} Eq.\ (\ref{eq:energydensity}).
This has the effect of raising the freeze-out temperature for the
processes of Eq.\ (\ref{eq:betaprocesses}) and therefore also the
instantaneous neutron-to-proton ratio, affecting again the value
of $Y_p$.  An interesting nucleosynthesis scenario that allows for
large neutrino asymmetries while simultaneously preserving the
standard primordial $^4{\rm He}$ yield is known as degenerate BBN
(DBBN), whereby an increased $n_n/n_p$ due to a large
$\xi_{\nu_{\mu}}$ and/or $\xi_{\nu_{\tau}}$ is compensated for by
a sizeable positive $\xi_{\nu_e}$ \cite{bib:dbbn}.  Several recent
combined analyses of (D)BBN and CMBR have generated the
constraints
\begin{equation}
\label{eq:constraints} -0.01 < \xi_{\nu_e} < 0.22, \qquad
|\xi_{\nu_{\mu},\nu_{\tau}}| < 2.6,
\end{equation}
assuming no neutrino oscillations \cite{bib:hansen,bib:kneller}.

\subsection{Neutrino oscillations?}

Naturally, one is curious to find out how the numbers in Eq.\
(\ref{eq:constraints}) hold up in the presence of oscillations. To
date, there are three pieces of evidence for neutrino
oscillations.  (i) Electron neutrinos originating from nuclear
reactions in the sun have been observed to transform largely into
$\nu_{\mu}$ and/or $\nu_{\tau}$ \cite{bib:solar}. Currently, these
oscillation parameters all provide acceptable explanations
for the $\nu_e$ deficit \cite{bib:solarfits}:
\begin{eqnarray}
\label{eq:solarparameters}
{\rm LMA} & \qquad & \delta m^2 \simeq 5 \times 10^{-5} \ {\rm
eV}^2 \qquad  \sin 2 \theta \simeq 0.8, \nonumber \\ {\rm LOW} &
\qquad & \delta m^2 \simeq  10^{-7} \ {\rm eV}^2 \qquad \quad \;
\; \sin 2 \theta \simeq {\rm large}, \nonumber \\ {\rm Vacuum} &
\qquad & \delta m^2 \simeq 10^{-10} \ {\rm eV}^2 \qquad \quad \;
\sin 2 \theta \simeq {\rm large}, \nonumber \\ {\rm SMA} & \qquad
& \delta m^2 \simeq 7 \times 10^{-6} \ {\rm eV}^2 \qquad \sin 2
\theta \simeq 0.05.
\end{eqnarray}
The LMA solution, however, is unanimously the most favoured. (ii)
The disappearance of atmospheric $\nu_{\mu}$ and $\bar{\nu}_{\mu}$
is generally attributed to oscillations into $\nu_{\tau}$ and
$\bar{\nu}_{\tau}$, with oscillation parameters $\delta m^2 \simeq
3 \times 10^{-3} \ {\rm eV}^2$ and $\sin 2 \theta \simeq 1$
\cite{bib:atmospheric}. (iii) The Liquid Scintillator Neutrino
Detector (LSND) experiment has reported evidence for $\bar{\nu}_{\mu}
\leftrightarrow \bar{\nu}_e$ oscillations with $\delta m^2
> 1 \ {\rm eV}^2$ and $\sin 2 \theta \simeq 0.03 \to 0.1$
\cite{bib:lsnd}. Since it is impossible to accommodate all three
signals with only three neutrinos, the most common practice is to
either introduce a fourth sterile neutrino (which in fact does
not provide a very good fit to the data \cite{bib:valle}),
or ignore the LSND result. We shall be considering the
latter option for simplicity.

In a rudimentary study by Lunardini and Smirnov
\cite{bib:lunardini}, it was suggested that three flavour
oscillations in a near-bimaximal mixing framework (i.e., if LMA
holds) necessarily lead to flavour equilibrium prior to BBN
(see also an earlier work by Savage, Malaney, and Fuller \cite{bib:savage}).
 This proposal was later examined in detail in the
numerical studies of Dolgov {\it et al.} \cite{bib:dhpprs}, and
then confirmed by analytical means in a paper by myself
\cite{bib:yyyw}, and in another independent work by Abazajian,
Beacom, and Bell \cite{bib:kev}. The common conclusion of these
works is that constraints from BBN on the $\nu_e$ asymmetry  must
now apply also to asymmetries in the $\nu_{\mu}$ and the
$\nu_{\tau}$ sectors, which constitutes a significant improvement
over the DBBN/CMBR bounds quoted in Eq.\ (\ref{eq:constraints}).

A rigorous three flavour analytical study is at present beyond our
means.  Fortunately, the hierarchical nature of the mass
splittings allows us to understand the full equilibration
mechanism in terms of two separate equilibrating transitions---the
first between $L_{\nu_{\mu}}$ and $L_{\nu_{\tau}}$, governed by
the atmospheric neutrino oscillation parameters, and the second
between $L_{\nu_e}$ and $L_{\nu_x}$, where $\nu_x$ denotes some
linear combination of $\nu_{\mu}$ and $\nu_{\tau}$, and the $\nu_e
\leftrightarrow \nu_x$ transformation proceeds with the
parameters of Eq.\ (\ref{eq:solarparameters}).\footnote{In principle, if the mixing angle
commonly known as $\theta_{13}$ is nonzero, a third equilibrating
transition is also present.  I shall not talk about this
possibility here.  The interested reader is referred to the
original works for details.}  I shall outline the mechanics of
these transformations.

\subsection{Equilibration of $L_{\nu_e}$ with $L_{\nu_x}$}

Suppose there is some pre-existing asymmetry in either or both of
the $\nu_e$ and the $\nu_x$ sectors.  The one-body reduced density
matrices for neutrinos and antineutrinos are as defined
previously.  These are again parameterised in terms of two
polarisation vectors ${\bf P}$ and $\bar{\bf P}$, and for the
present application, we choose the normalisation factor $N_0(p)$
to be $N_{\rm eq}(0)$ as defined in Eq.\ (\ref{eq:fermidirac}).
The evolution of the neutrino--antineutrino ensemble at
temperatures $T \sim {\cal O} (1) \to {\cal O}(10) \ {\rm MeV}$ is
governed by Eq.\ (\ref{eq:anti}), where the volume $V$
accompanying the self interaction term is already absorbed into
the present definition of $N_0(p)$, and the background potentials are
\begin{equation}
A_{CP+}  =  - \frac{8 \sqrt{2} G_F p}{3 m^2_W} E_{ee}, \qquad
A_{CP-} = \sqrt{2} G_F n_{\gamma} \eta,
\end{equation}
in which $E_{ee}$ is the electron--positron energy density, $m_W$ is
the mass of the $W$ boson, and $\eta$ is the electron--positron
asymmetry. By universal charge neutrality, $\eta$ must be of
the order of the baryon asymmetry so that $A_{CP-}$ is
negligible in comparison with $A_{CP+}$. The quantity we are
interested in is the difference between the two asymmetries
\begin{eqnarray}
\label{eq:le-lx} L_{\nu_e} - L_{\nu_x}  &=& \frac{1}{n_{\gamma}}
\int \left[(N_{\nu_e} - N_{\bar{\nu}_e}) - (N_{\nu_{x}} -
N_{\bar{\nu}_{x}}) \right]  = \frac{1}{ n_{\gamma}} \int (P_z -
\bar{P}_z ) N_{\rm eq}(0)   \nonumber
\\ &=& \frac{1}{n_{\gamma}} (J_z-\bar{J}_z).
\end{eqnarray}
Hence we would like to track the evolution of the vector ${\bf I}
\equiv ({\bf J} - \bar{\bf J})/n_{\gamma}$ (note the new
definition), and flavour equilibrium necessarily implies $I_z
=0$.

The next question is, are all the modes synchronised so that we
can apply results from the previous sections?  The answer comes
from figuring out the conditions under which Eq.\ (\ref{eq:synchi})
holds true, and for the temperature range of interest, this
amounts to requiring only that the disparity between the initial
asymmetries be larger than, say, $10^{-5}$ in magnitude
\cite{bib:yyyw}. Then,
assuming that all neutrinos and antineutrinos are initially in
thermal and chemical equilibrium, i.e.,
\begin{equation}
{\bf P}^{\rm i} \cdot \hat{\bf I}^{\rm i} \simeq \frac{N_{\rm
eq}(\xi_{\nu_e}^{\rm i}) - N_{\rm eq}(\xi_{\nu_x}^{\rm i})}{N_{\rm
eq}(0)}, \qquad \bar{\bf P}^{\rm i} \cdot \hat{\bf I}^{\rm i}
\simeq \frac{N_{\rm eq}(-\xi_{\nu_e}^{\rm i}) - N_{\rm
eq}(-\xi_{\nu_x}^{\rm i})}{N_{\rm eq}(0)},
\end{equation}
the general compound evolution equation (\ref{eq:compound}) can be
rewritten explicitly for this application as
\begin{equation}
\label{eq:didtfinal}
 \partial_t {\bf I} \simeq \frac{3}{2}\frac{\widetilde{y} \
  ({\xi^{\rm i}_{\nu_e}}^2 -
{\xi^{\rm i}_{\nu_{x}}}^2 )}{| \pi^2 (\xi^{\rm i}_{\nu_e}
-\xi^{\rm i}_{\nu_{x}}) +  ({\xi^{\rm i}_{\nu_e}}^3 - {\xi^{\rm
I}_{\nu_{x}}}^3) |}  \left( \frac{\Delta m^2}{2 \widetilde{p}}
{\bf B} - \frac{8 \sqrt{2} G_F \widetilde{p}}{3 m^2_W} E_{ee} {\bf
z} \right) \times {\bf I},
\end{equation}
where
\begin{equation}
\label{eq:tildep}
 \widetilde{y} = \frac{\widetilde{p}}{T}
\equiv \sqrt{\pi^2 + \frac{1}{2} \left( {\xi^{\rm i}_{\nu_e}}^2 +
{\xi^{\rm i}_{\nu_{x}}}^2 \right)}
\end{equation}
represents some average momentum.

Equation  (\ref{eq:didtfinal}) has a straightforward
interpretation.  Consider the terms inside the parentheses. These
are identically the vacuum and electron--positron background terms
that one would find in a single momentum evolution equation, and
control the ensemble's synchronised mixing angle,
\begin{equation}
\sin 2 \theta_{\rm synch}  =  \left. \frac{\sin 2 \theta}{\sqrt{(
2p A_{CP+} /\delta m^2  - \cos 2 \theta)^2 + \sin^2 2 \theta}}
\right|_{p = \widetilde{p}},
\end{equation}
evaluated for $A_{CP+} =  - 8 \sqrt{2} G_F \widetilde{p}E_{ee}/3
m^2_W$ at $p = \widetilde{p}$. The factor multiplying the
bracketed terms, which we shall label as $\kappa$,
\begin{equation}
\label{eq:kappa} \kappa = \frac{3}{2}\frac{\widetilde{y} \
({\xi^{\rm i}_{\nu_e}}^2 - {\xi^{\rm i}_{\nu_{x}}}^2 )}{| \pi^2
(\xi^{\rm i}_{\nu_e} -\xi^{\rm i}_{\nu_{x}}) +  ({\xi^{\rm
i}_{\nu_e}}^3 - {\xi^{\rm i}_{\nu_{x}}}^3) |},
\end{equation}
does not play a role in determining the effective mixing angle. It
does, however, influence directly the oscillation frequency,
\begin{equation}
\frac{\delta m^2_{\rm synch}}{2 \widetilde{p}} = \left. \kappa
\delta m^2 \sqrt{(2p A_{CP+}/\delta m^2  - \cos 2 \theta)^2 +
\sin^2 2 \theta} \right|_{p=\widetilde{p}}.
\end{equation}
For instance, $\delta m^2_{\rm synch}$ vanishes for
$\xi_{\nu_e}^{\rm i} = - \xi_{\nu_x}^{\rm i}$, and oscillations
are switched off completely.

An approximate solution to Eq.\ (\ref{eq:didtfinal}) can be
obtained in the adiabatic limit.  Specifically, the expression for
the variable $I_z$ as a function of time is
\begin{equation}
\label{eq:Iz}  I_z  \equiv L_{\nu_e} - L_{\nu_x}\simeq \left( \cos
2 \theta_c \cos 2 \theta_c^{\rm i} + \sin 2 \theta_c \sin 2
\theta^{\rm i}_c \ \cos
 \int^t_{t_{\rm i}}  \frac{\Delta m^2_{\rm eff}}{2
\widetilde p} dt' \right) I_z^{\rm i},
\end{equation}
assuming the validity of the adiabatic condition
\begin{equation}
\label{eq:adparameter} \gamma \equiv \left|\frac{V_z \
\partial_t V_x - V_x \ \partial_t V_z}{\kappa (V_x^2 +
V_z^2)^{3/2}} \right|_{p=\widetilde{p}} < 1,
\end{equation}
where $V_x = (\delta m^2/2p) \sin 2 \theta$, and $V_z = A_{CP+} -
(\delta m^2/2p) \cos 2 \theta$, both evaluated for $p = \widetilde{p}$.
Equation (\ref{eq:Iz}) predicts for maximal vacuum mixing an
MSW-like effect, transforming $I_z$ from $I_z^{\rm i}$ to $0$
(plus some small amplitude oscillations) when vacuum oscillations
overcome refractive matter effects.  The temperature at which this
equilibrating transition takes place can be established roughly by solving
\begin{equation}
\frac{8 \sqrt{2} G_F \widetilde{p}}{3 m_W^2}  \simeq
\frac{|\delta m^2|}{2 \widetilde{p}}.
\end{equation}
For initial chemical potentials satisfying the constraints
(\ref{eq:constraints}), the equilibration temperature
turns out to be $\simeq 2.6\ {\rm
MeV}$ for the LMA solution. In the cases of the LOW and the Vacuum
mass splittings, the  temperatures are $\simeq0.9 \ {\rm MeV}$ and
$\simeq 0.3\ {\rm MeV}$ respectively. Evidently, only the LMA
equilibrating transition can take place well ahead of BBN.\footnote{The real LMA
solution encompasses a range of mixing parameters that are merely
large, but not maximal \cite{bib:solarfits}.
Thus only a partial equilibrium between
the asymmetries can be achieved from synchronised oscillations
alone. However, collisions with the background medium (i.e.,
momentum-changing non-forward scattering) are inevitable, and
these are expected to drive the equilibration to a more complete
state.}

For the oscillation parameters of the SMA solution,  $I_z$ remains
close to its initial value even after the ``transition'' at $T
\simeq 1.9\ {\rm MeV}$, since both the vacuum and the
background terms are predominantly in the negative $z$-direction
for $\delta m^2
> 0$, and the usual MSW resonance condition
cannot be satisfied.

A second deciding factor on the efficacy of flavour equilibration
is the adiabaticity of the transition from matter-suppressed to
vacuum oscillations. Unlike that encountered in, for instance,
solar neutrino analyses, the adiabaticity parameter $\gamma$ of
Eq.\ (\ref{eq:adparameter}) is strongly dependent on the initial
conditions. Take for concreteness the case of $\xi_{\nu_e}^{\rm
i}=0$.   The adiabatic condition $\gamma < 1$  always holds for
LMA if $|\xi_{\nu_x}^{\rm i}|
> 0.01$, but can be badly violated at the transition point for the
LOW $\delta m^2$ unless $|\xi_{\nu_x}^{\rm i}| > 0.1$ \cite{bib:yyyw}.
In the case
of maximal mixing,  violation of adiabaticity at the transition
point generally results in large amplitude ``post-transition''
oscillations about the equilibrium point at an angular frequency
roughly equal to $\delta m^2_{\rm synch}/2 \widetilde{p}$.
Naturally, this is quite a separate phenomenon from true
equilibration.   On the other hand, an adiabaticity parameter that
evaluates to infinity at all times (because, for example,
$\kappa\to0$) signifies that there is no transition at all.  From
the perspective of equilibrating two vastly different asymmetries,
the requirement of $|\xi^{\rm i}_{\nu_x}| > 0.01$ (assuming
$\xi^{\rm i}_{\nu_e} = 0$) in the LMA case for a smooth transition
is, by definition, not a major concern.

\subsection{Equilibration of $L_{\nu_{\mu}}$ and
$L_{\nu_{\tau}}$, and new constraints on neutrino--antineutrino
asymmetries}

Super-Kamiokande's atmospheric neutrino data indicate maximal
mixing between $\nu_{\mu}$ and $\nu_{\tau}$, with a mass splitting
that is some two orders of magnitude larger than the maximum
acceptable solar $\delta m^2$. This means, in general terms, that
equilibration of the asymmetries in the $\nu_{\mu}$ and the
$\nu_{\tau}$ sectors will occur in a manner similar to that
outlined before, but at a higher temperature, and subject to a
background potential proportional to the muon--antimuon energy
density $E_{\mu \mu}$.  A  comparison of the magnitudes of the
vacuum and the background potential terms returns an equilibration
temperature of $\simeq 12 \ {\rm MeV}$ for $\delta m^2 = 3 \times
10^{-3} \ {\rm eV}^2$ and $p = \widetilde{p} \simeq \pi/T$, and
complete equilibrium between $L_{\nu_{\mu}}$ and $L_{\nu_{\tau}}$
can always be achieved because of maximal mixing and
collision-induced flavour relaxation not discussed here
\cite{bib:dhpprs,bib:yyyw,bib:kev}.

Thus, summing up the findings, we conclude that all active
neutrino--antineutrino asymmetries or chemical potentials must
come to an equal value before the onset of BBN, if the flavour
oscillation parameters are those indicated by the atmospheric
neutrino data and by the solar LMA solution.   The stringent
constraint imposed by the primordial $^4{\rm He}$ abundance on
$\xi_{\nu_e}$,
\begin{equation}
\xi_{\nu_e} \simeq \xi_{\nu_{\mu}} \simeq \xi_{\nu_{\tau}} < 0.07,
\end{equation}
now applies to all three neutrino flavours.

\section{Application 2: $r$-process
nucleosynthesis in a supernova hot bubble}

\label{sec:section3}

Heavy nuclei with mass number $A > 70$ are predominantly produced
by slow and rapid neutron capture---the $s$- and the $r$-processes.
The most plausible site for $r$-process nucleosynthesis suggested
so far is the hot bubble between a protoneutron star and the
escaping shock wave in a core collapse supernova a few seconds
after core bounce.  A key condition for this process to be
successful, however, is that the environment must be rich in
neutrons (relative to protons), and neutrinos released from the
cooling of the nascent neutron star play an important role in establishing
this condition.

Just as in the early universe, the neutron-to-proton ratio is
fixed by the $\beta$-processes in Eq.\ (\ref{eq:betaprocesses}).
Assuming charge neutrality, it is common to express this ratio in
terms of the electron fraction $Y_e$:
\begin{equation}
\frac{n_n}{n_p} = \frac{1}{Y_e} -1.
\end{equation}
Clearly, $Y_e < 0.5$ will satisfy the minimum requirement for
neutron predominance, but a successful $r$-process may call for $Y_e
<0.45$.  Because neutrinos are far more abundant than
electrons and positrons, $Y_e$ is governed largely by the
neutrino spectra and fluxes.   Near the weak freeze-out radius
($\sim 30 \to 35 \ {\rm km}$), this is well approximated by
\cite{bib:qian}
\begin{equation}
\label{eq:electronfraction}
Y_e \simeq \left( 1 +
\frac{L_{\bar{\nu}_e} \bar{\epsilon}}{L_{\nu_e} \epsilon}
\right)^{-1},
\end{equation}
where $L_{\nu_e}$ and $L_{\bar{\nu}_e}$ are the electron neutrino
and antineutrino luminosities (typically of order $10^{51}\ {\rm
erg \ s}^{-1}$), $\epsilon \equiv \langle E^2_{\nu_e} \rangle/
\langle E_{\nu_e} \rangle$, and $\bar{\epsilon} \equiv \langle
E^2_{\bar{\nu}_e} \rangle/ \langle E_{\bar{\nu}_e} \rangle$.
Neutrinos streaming out of the neutrino sphere
 at a radius of $\sim 10 \ {\rm km}$ generally
possess spectra of Fermi--Dirac form, with mean energies
\begin{equation}
\langle E_{\nu_e} \rangle \simeq 11 \ {\rm MeV}, \qquad \langle
E_{\bar{\nu}_e} \rangle \simeq 16 {\rm MeV}, \qquad \langle
E_{\nu_{\mu}, \bar{\nu}_{\mu},\nu_{\tau}, \bar{\nu}_{\tau}}
\rangle \simeq 25 \ {\rm MeV}.
\end{equation}
These numbers yield for $Y_e$ a value of $\sim 0.41$, allowing
for a successful $r$-process, in the absence of neutrino
oscillations.

Contrastingly,  if flavour oscillations between $\nu_e$ and
$\nu_{\mu,\tau}$ occur outside the neutrino sphere but still
within the weak freeze-out radius,
then the two dissimilar neutrino spectra will be exchanged to
various degrees depending on the oscillation parameters, and $Y_e$
becomes modified as a generic consequence \cite{bib:qian}. Since a
large, positive electron--positron asymmetry is always present in  the
background medium, and the interaction Hamiltonian  dominated by the
$CP$ asymmetric potential $\sqrt{2} G_F (n_{e^-} - n_{e^+})$,
conventional wisdom tells us that the MSW resonance condition
cannot be satisfied simultaneously by both neutrinos and
antineutrinos. If the former meets the condition, the more
energetic $\nu_{\mu}$'s will be transformed across the resonance
into $\nu_e$'s, while the $\bar{\nu}_e$ energy spectrum remains
virtually unaffected.  This then leads to an over-production of
electrons, pushes $Y_e$ to above $0.5$, and consequently ruins the
$r$-process. The mass splitting required for such oscillations to
be effective is $\delta m^2 > 1 {\rm eV}^2$, which overlaps with
the region of parameter space needed to explain the LSND result
\cite{bib:lsnd}.

This situation changes drastically when neutrino--neutrino forward
scattering is also taken into account.  Specifically, for the
nominal neutrino luminosity $L_{\nu} = 10^{51} \ {\rm erg}$, the
electron fraction always stays below $0.5$, such that the minimum
requirement for a neutron-rich environment can always be
satisfied, irrespective of neutrino oscillations.  This
interesting result can be easily understood in terms of
synchronisation \cite{bib:sn}.

As noted before, when neutrino--neutrino forward scattering
constitutes the dominant interaction potential, it synchronises
both neutrinos and antineutrinos, regardless of the nature of the
non-neutrino background medium.  Consequently, if the neutrinos
were to encounter an MSW resonance, the antineutrinos would also
be dragged along for the ride, and any spectral swap that
accompanies the transitions would occur with identical efficiency.
The net effect is that the post-resonance $\nu_e$ and
$\bar{\nu}_e$ will now have identical spectra, having taken over those
which once belonged to $\nu_{\mu, \tau}$ and $\bar{\nu}_{\mu,
\tau}$ respectively.  Assuming that $L_{\nu_e} \simeq
L_{\bar{\nu}_e}$, Eq.\ (\ref{eq:electronfraction}) always gives $Y_e
\simeq 0.5$ for the electron fraction.

\section{Conclusions}

I have reported in this talk the findings of a series of recent
works devoted to explaining and applying the phenomenon of
synchronised flavour oscillations due to neutrino self
interaction, in the early universe and in a core collapse
supernova.

For the ``explanation''  part, I presented the original
interpretation of Pastor, Raffelt, and Semikoz \cite{bib:synch},
as well as extensions to it introduced by myself \cite{bib:yyyw},
and separately, by Abazajian, Beacom, and Bell \cite{bib:kev}.

Application-wise, the analyses of Dolgov {\it et al.}, and Refs.\
\cite{bib:yyyw} and \cite{bib:kev} demonstrate that neutrino self
interactions in the epoch immediately prior to BBN  are
sufficiently intense to synchronise the multi-momentum flavour
oscillations, which, together with the large mixing angles indicated by present
neutrino data, allows for the efficient equilibration of
all pre-existing neutrino--antineutrino asymmetries.
 I have reviewed these works
in considerable detail, and relayed the unanimous conclusion that
stringent BBN bounds on the electron neutrino--antineutrino
asymmetry must now apply also to asymmetries in the $\nu_{\mu}$
and the $\nu_{\tau}$ sectors.

Lastly, I spoke briefly about a possible upset of the neutron-rich
environment and thus the success of $r$-process nucleosynthesis in
a supernova hot bubble by ``conventional'' neutrino oscillations.
I proceeded to report the new results of Pastor and Raffelt \cite{bib:sn},
which show that, if neutrino self interactions are properly taken
into account, synchronised oscillations  will always prevent the
neutron-to-proton ratio from dropping below the minimum value
required for neutron predominance.

And on this note, I rest my case.

\section*{Acknowledgments}

I wish to thank the workshop organisers for their support.  This
work was supported also by the U.~S. Department of Energy under
grant DE-FG02-84ER40163.

\end{document}